# From the Coulomb breakup of halo nuclei to neutron radiative capture

**P. Capel*  and Y. Nollet**
*Physique Nucléaire et Physique Quantique (CP 229)*
*Université Libre de Bruxelles (ULB)*
*B-1150 Brussels, Belgium*
*E-mail:* `pierre.capel@ulb.ac.be`

Coulomb breakup is used to infer radiative-capture cross sections at astrophysical energies. We test theoretically the accuracy of this indirect technique in the particular case of $^{15}$C, for which both the Coulomb breakup to $^{14}$C+n and the radiative capture $^{14}$C(n,$\gamma$)$^{15}$C have been measured. We analyse the dependance of Coulomb-breakup calculations on the projectile description in both its initial bound state and its continuum. Our calculations depend not only on the Asymptotic Normalisation Coefficient (ANC) of the $^{15}$C ground state, but also on the $^{14}$C-n continuum. This questions the method proposed by Summers and Nunes [Phys. Rev. C **78**, 011601 (2008), ibid. **78**, 069908 (2008)], which assumes that an ANC can be directly extracted from the comparison of calculations to breakup data. Fortunately, the sensitivity to the continuum description can be absorbed in a normalisation constant obtained by a simple $\chi^2$ fit of our calculations to the measurements. By restricting this fit to low $^{14}$C-n energy in the continuum, we can achieve a better agreement between the radiative-capture cross sections inferred from the Coulomb-breakup method and the exact ones. This result revives the Coulomb-breakup technique to infer neutron radiative-capture capture to loosely-bound states, which would be very useful for *r*- and *s*-process modelling in explosive stellar environments.

*Speaker.

## 1. Introduction

Radiative captures are reactions during which two nuclei *b* and *c* fuse to form a nucleus *a* by emiting a $\gamma$:

$$b+c \rightarrow a+\gamma. \tag{1.1}$$

These reactions, also noted $c(b,\gamma)a$, take place in many astrophysical sites. For example, as discussed by Mossa during this conference [1], $d(p,\gamma)^3$He is one of the reactions of the pp chain which takes place in the Sun and has happened during the Big-Bang nucleosynthesis. The *s* and *r* processes that take place during supernova-explosions consist of sequences of neutron radiative captures $(n,\gamma)$ [2, 3].

To constrain astrophysical models, the corresponding cross sections need to be measured at the relevant energies. These energies being rather low (of the order of a few tens of keV in stars), the direct measurement of the radiative-capture cross sections can be quite difficult, either because of the Coulomb barrier between the colliding nuclei or because they involve neutrons. The former hinders the capture by repelling the colliding nuclei from each other, and the seconds are difficult to handle experimentally. Hence the interest in indirect techniques to infer these cross sections [4, 5, 6]. Coulomb breakup is one of them [4, 5]. In that reaction, the projectile dissociates into lighter fragments through its interaction with a heavy (high *Z*) target *T*:

$$a+T \rightarrow b+c+T. \tag{1.2}$$

Assuming the dissociation to be due to the sole Coulomb interaction, the reaction can be described as an exchange of virtual photons between the projectile and the target. It can thus be seen as the time-reversed reaction of the radiative capture of the fragments, which should enable us to deduce easily the radiative-capture cross section from breakup measurements [4, 5].

Using accurate reaction models, various studies have shown that higher-order effects and other reaction artefacts play a significant role in Coulomb breakup, which hinder the simple extraction of radiative-capture cross sections from breakup measurements [7, 8]. The case of $^{15}$C is of particular interest to analyse this indirect method as both its Coulomb breakup [9] and the radiative capture $^{14}$C$(n,\gamma)^{15}$C [10] have been measured accurately.

Summers and Nunes have recently proposed an innovating analysis of the Coulomb-breakup measurement [11]. They have confirmed the significant influence of dynamical effects observed in previous analyses [7, 8] and, accordingly, the need of an accurate reaction model to study properly such reactions. Since breakup reactions are mostly peripheral, in the sense that they probe mostly the tail of the projectile wave function [12], they have suggested to use the comparison between their calculations and the measurements to deduce the Asymptotic Normalisation Coefficient (ANC) of the $^{15}$C bound state from experimental data [11]. They then suggest to rely on this inferred ANC to compute the cross section for the $^{14}$C$(n,\gamma)^{15}$C radiative capture. Their idea leads to inferred radiative-capture cross sections in good agreement with the direct measurements [11, 13].

In the present work we analyse the influence of the description of the $^{14}$C-n continuum upon breakup calculations. In Sec. 2, we present the model of $^{15}$C we use in this study, emphasising on the $^{14}$C-n interaction used in the continuum. We then discuss our results obtained for the Coulomb breakup of $^{15}$C on Pb at 68$A$ MeV (Sec. 3) and present our analysis of the extraction of the cross section for the radiative capture $^{14}$C(n,$\gamma$)$^{15}$C following the prescription of Summers and Nunes in Sec. 4. We show also how this method can be improved by selecting the data at low energy in the $^{14}$C-n continuum. We end by the conclusions and perspectives of this work.

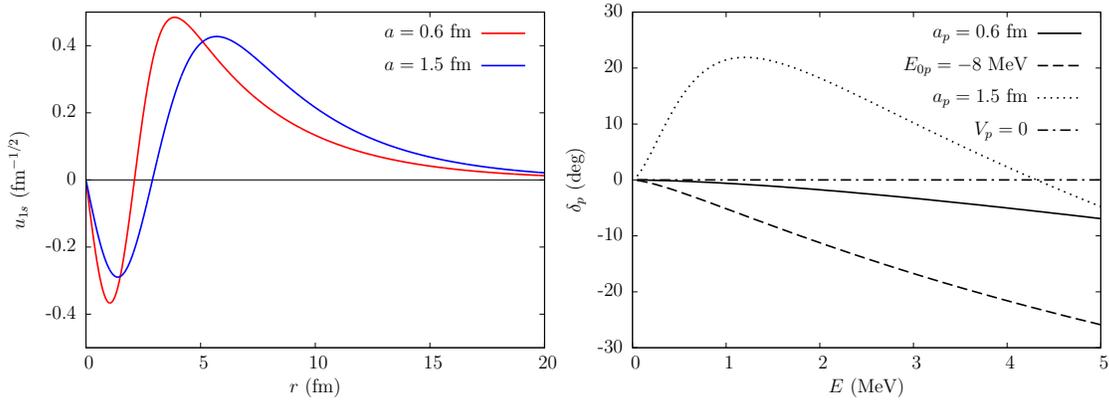

**Figure 1:** Left: wave functions describing the $^{15}$C $s$ ground state using a regular diffuseness ($a = 0.6$ fm, red) and a very diffuse potential ($a = 1.6$ fm, blue). Right: $p$-wave phaseshifts in the $^{14}$C-n continuum obtained with four different potentials (see text for details).

## 2. Model of $^{15}$C

We consider a simple two-body model of $^{15}$C, viz. a neutron loosely bound to an inert $^{14}$C core. The $^{14}$C-n interaction is simulated by a phenomenological Woods-Saxon potential, whose depth is adjusted to reproduce the low-energy spectrum of $^{15}$C. For simplicity, the spins of the fragments are ignored in this study. Hence the $1/2^+$ ground state is described as the $1s$ state of the potential while the $5/2^+$ excited bound state is reproduced by the $0d$ state. To confirm the sensitivity of the breakup calculations to the ANC of the bound states, we consider Woods-Saxon form factors with two different geometries: one with a regular diffuseness ($a = 0.6$ fm) leading to a regular ANC, and one with a large diffuseness ($a = 1.5$ fm), which leads to a much larger ANC, i.e. with a density of probability significantly shifted towards large $^{14}$C-n distances [see Fig. 1 (left)].

In this way, the $^{14}$C-n potential is constrained in the $s$ and $d$ waves, but not in the $p$ wave. Since the dominant transition, in both the Coulomb breakup of $^{15}$C and the radiative capture $^{14}$C(n,$\gamma$)$^{15}$C, is an E1 transition to (resp. from) the $p$ continuum from (resp. to) the $s$ ground state, the description of the $p$ wave is the one that matters, if the description of the continuum plays any role in those reactions. To test this hypothesis, we use four different potentials in the $p$ wave in order to vary the phaseshift as much as possible [see Fig. 1 (right)]. The first one is the regular $^{14}$C-n potential used to describe $^{15}$C ground state ($a_p = 0.6$ fm, solid line). The second exhibit the same geometry but has its depth adjusted to set the forbidden $0p$ bound state at the one-neutron separation energy of $^{14}$C ($E_{0p} = -8$ MeV, dashed line). To obtain a significant change, we also use the very diffuse potential mentioned earlier ($a_p = 1.5$ fm, dotted line). Finally, we also test what happens when the $^{14}$C and the neutron do not interact in the $p$ wave ($V_p = 0$, dash-dotted line).

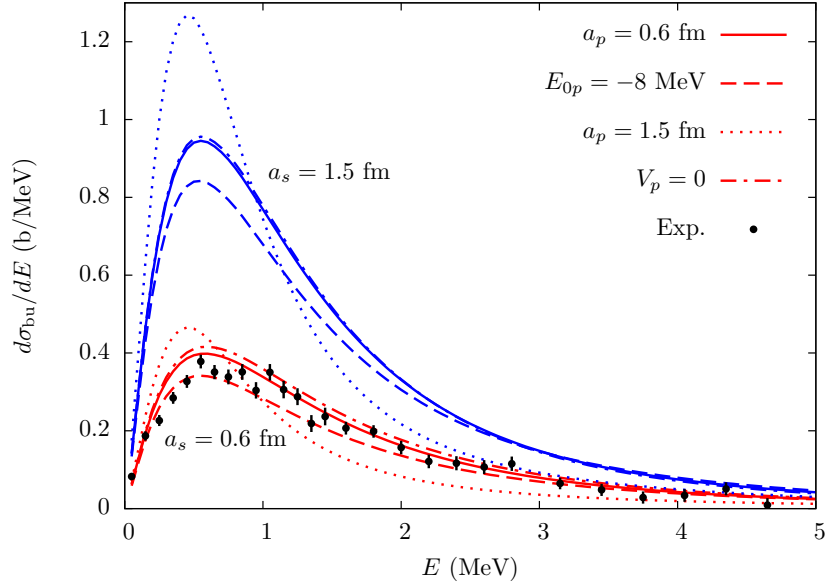

**Figure 2:** Breakup cross section of $^{15}$C on Pb at 68$A$ MeV as a function of the $^{14}$C-n relative energy $E$ after dissociation. The calculations are performed with the $^{14}$C-n potentials described in Sec. 2. Data are from Ref. [9].

## 3. Coulomb breakup of $^{15}$C on Pb at 68$A$ MeV

Combining the two descriptions of $^{15}$C in its ground state and the four ones of its $p$ continuum, we obtain eight models of the projectile, which we use to compute the Coulomb breakup of $^{15}$C on Pb at 68$A$ MeV, corresponding to the RIKEN experiment [9]. The calculations are performed with the Dynamical Eikonal Approximation, which provides excellent agreement with experiments on one-nucleon halo nuclei [14, 15]. Fig. 2 shows the corresponding $^{15}$C breakup cross sections as a function of the $^{14}$C-n relative energy $E$ after dissociation. As Summers and Nunes [11], we observe a significant influence of the ground-state ANC. The cross sections computed with the diffuse state [blue (top) curves] are much larger than those obtained with the regular diffuseness [red (bottom) curves]. However, the ANC is not the only parameter that affects the calculations. As already observed in Ref. [16], the description of the $p$ continuum significantly influences the value of the breakup cross section. Interestingly, the order of the curves is the same for both descriptions of the ground state. The calculations performed setting the 0$p$ forbidden bound state at the $^{14}$C one-neutron separation energy (dashed lines) are about 15% below the calculations performed using the regular $s$-wave potential in the $p$ wave (solid lines). The latter is very close to the results obtained with no $^{14}$C-n interaction in the continuum, as expected from the fact that both descriptions lead to very similar phaseshifts (see Fig. 1 right). Using the very diffuse potential in the $p$ wave (dotted lines) leads to a significant change of the shape of the breakup cross section compared to the other calculations. This is to be related to the weird behaviour of the $p$ phaseshift observed for that potential in Fig. 1. These results put at stake the idea of Summers and Nunes. The significant influence of the $^{14}$C-n continuum observed in Fig. 2 shows that the ANC extracted by fitting breakup calculations to experimental data is spoiled by the description of the continuum.

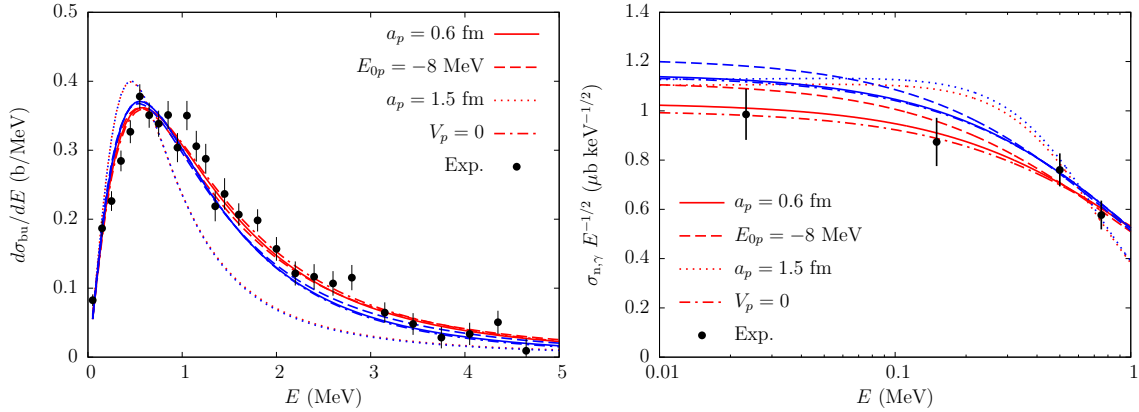

**Figure 3:** Left: theoretical breakup cross sections scaled to the data of Ref. [9]. Right: radiative-capture cross sections obtained with our eight descriptions of $^{15}$C scaled with the factors obtained from the fit of the breakup data. Experimental radiative-capture data are from Ref. [10].

## 4. Inferring the cross section for the radiative capture $^{14}$C(n,$\gamma$)$^{15}$C

Following the results presented in the previous section, is the method proposed in Ref. [11] to extract the radiative capture cross section from breakup reliable? To answer this question, we follow the method suggested by Summers and Nunes [11] and compare the prediction for the radiative-capture cross section obtained by the fit of our eight theoretical breakup cross sections to the RIKEN data. This fit consist simply in multiplying the theoretical cross section by a scaling factor and adjust that factor to obtain the best $\chi^2$ with the data. That factor is then used in the direct radiative capture calculation.

The results of this fit are displayed in Fig. 3 (left). We observe a good agreement between the fitted theoretical cross sections and the data, but for the cases in which the (very unrealistic) diffuse potential is used in the $^{14}$C-n $p$ wave (dotted blue and red lines). In these two cases no scaling parameter could enable us to reproduce the data with such a narrow peak.

Originally this method is based on the idea that the scaling factor would correct the theoretical ANC to its actual value, and that this value can then be used to predict reliably the radiative-capture cross section. However, the influence of the continuum description on the calculations tells us that no actual ANC can be extracted in such a way. Nevertheless, we naively apply this method to our calculations. The corresponding radiative-capture cross sections are depicted in Fig. 3 (right) as a function of the $^{14}$C-n energy. They are found in fairly good agreement with the direct data of Reifart *et al.* [10]. We observe a larger spread of our theoretical predictions than that obtained by Summers and Nunes [11], and an average value slightly larger than the direct data. Nevertheless, this good result indicates that most of the dependence of our calculations to the projectile description, in both its bound and scattering states, can be captured in this scaling constant, and that this dependence is very similar for both breakup and radiative-capture processes. This is mostly due to the fact that both reactions are sensitive to the same physical inputs from the $^{15}$C structure, namely the ANC of its bound state and the phaseshift in its continuum. Thanks to this, the idea of Summers and Nunes provides reliable radiative-capture cross sections, even when varying the projectile description in the continuum. However, according to what has been said before, the scaling factor extracted from the $\chi^2$ fit of the the breakup calculation to the experimental data cannot be directly related to the ground-state ANC.

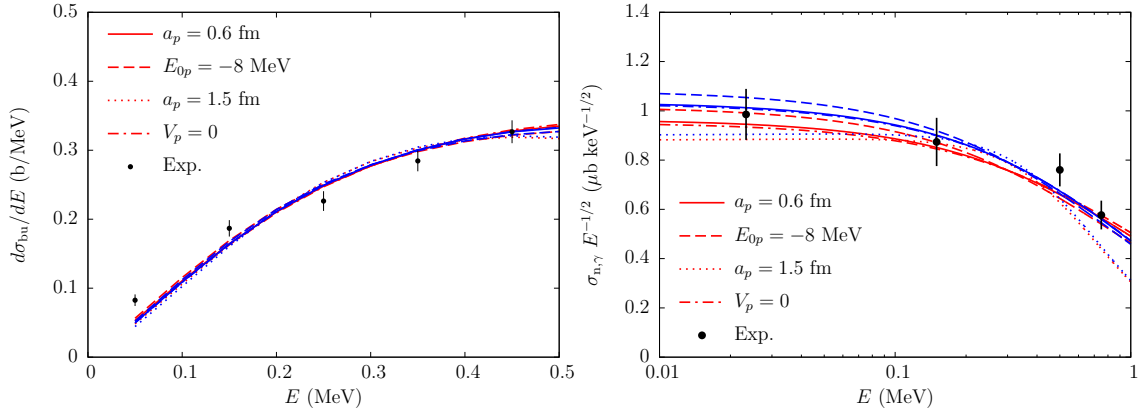

**Figure 4:** Left: theoretical breakup cross sections scaled to the data of Ref. [9] restricted in the low-energy range ($E < 0.5$n MeV). Right: radiative-capture cross sections obtained that low-energy fit.

Apart from the sensitivity of the breakup calculations to the projectile continuum, Fig. 3 (left) teaches us that all calculations, even the very unrealistic ones, exhibit the same behaviour at low energy, e.g. before the cross sections reach a maximum at about 0.5 MeV. This energy range is also the range of astrophysical interest: the radiative-capture cross sections are needed at low energy, not at a few MeV in the continuum. Hence the idea to apply Summers and Nunes' method not on the whole range of the breakup data, but to restrict it at low energy.

The result of such a fit is displayed in Fig. 4. The left panel shows that all breakup cross sections, including the very exotic ones, fall on top of each other when fitted below 0.5 MeV. Using the fitting parameter obtained in such a way, we get radiative-capture cross sections in much better agreement with the direct data, even with unrealistic $^{14}$C-n potentials [see Fig. 4 (right)]. This interesting result confirms the validity of the method proposed by Summers and Nunes and that it enables to capture most of the nuclear-structure inputs that matter in the radiative capture In addition, it shows that it works better if applied to low $^{14}$C-n continuum energies, which actually matter in astrophysical processes.

## 5. Conclusions and perspectives

Coulomb breakup has been proposed as an indirect technique to infer radiative-capture cross section at astrophysical energies [4, 5]. This idea is based on the hypothesis that Coulomb breakup, corresponding to an exchange of virtual photons between the projectile and the target, can be seen as the time-reversed reaction of the radiative capture. Unfortunately, the direct extraction of the latter cross section from the former's can only be done if the Coulomb breakup takes place at first order, which we know is not the case [7, 8]. To circumvent this issue, Summers and Nunes have proposed a method based on the fact that breakup reactions are mostly peripheral [12]. In this method, an ANC for the projectile bound state is extracted from the fit of accurate breakup calculations to breakup data [11, 13].

In this work, we have studied the sensitivity of this method to the description of the projectile continuum, which has been ignored in Summers and Nunes' analysis. From accurate calculations of the breakup of $^{15}$C on Pb at 68$A$ MeV using different descriptions of the projectile, we have shown that the sensitivity to the $^{14}$C-n continuum cannot be overlooked and that the scaling factor extracted from the fit suggested by Summers and Nunes contains information not only about the ANC of the projectile bound state, but also about its continuum. Nevertheless, the method works fine. We understand this by the fact that the structure information absorbed in this fitting procedure is important for both processes.

We have observed that this fit does not have much sense for the extreme descriptions of the $^{14}$C-n continuum, as they lead to significant distortions in the breakup cross sections compared to regular potentials. To account for this, we suggest to restrict the fit suggested by Summers and Nunes to low $^{14}$C-n energies, i.e. those that are of significance for astrophysics. The radiative-capture cross sections obtained in such a way are in excellent agreement with the direct data. This variant of Summers and Nunes' idea hence enables to extract reliably radiative-capture cross section from Coulomb breakup data without having to worry about the description of the two-body projectile in both its bound state and its continuum. This, in a sense, revives the original idea of Baur, Bertulani and Rebel [4]. It would be interesting to see if this variant can be improved by selecting breakup data at forward angles, where the process is fully dominated by the Coulomb interaction. Another interesting perspective is to see whether this method can be applied to charged cases, like for $^3$He($\alpha,\gamma$)$^7$Be or $^7$Be(p,$\gamma$)$^8$B.

## Acknowledgments

This work is part of the Belgian Research Initiative on eXotic nuclei (BriX), program P7/12 on inter-university attraction poles of the Belgian Federal Science Policy Office. It was supported in part by the Research Credit No. 19526092 of the Belgian Funds for Scientific Research F.R.S.-FNRS.